\documentclass{article}
\usepackage[margin=3cm]{geometry}

\usepackage[utf8]{inputenc} 
\usepackage[T1]{fontenc}    
\usepackage{url}            
\usepackage{booktabs}       
\usepackage{amsfonts}       
\usepackage{microtype}      
\usepackage[noadjust]{cite}

\usepackage{graphicx}
\usepackage{bm}
\usepackage{amsfonts}
\usepackage{xcolor}
\usepackage{array}
\usepackage{enumitem}
\usepackage{verbatim}
\usepackage[colorlinks = true,
            linkcolor = blue,
            urlcolor  = blue,
            citecolor = blue,
            anchorcolor = blue]{hyperref}

\usepackage{amsmath}    
\usepackage{amssymb}
\usepackage{caption}
\usepackage{subcaption}

\title{TURB-Smoke. A database of Lagrangian pollutants emitted from point‑sources and dispersed in turbulent flows}
\date{}

\author{
  Luca Biferale \\
  Department of Physics \& INFN, \\ 
  Tor Vergata University of Rome, \\ 
  Via della Ricerca Scientifica 1, 00133, Rome (Italy) \\
  \texttt{biferale@roma2.infn.it} \\
   \and
  Fabio Bonaccorso \\
  Department of Physics \& INFN, \\ 
  Tor Vergata University of Rome, \\ 
  Via della Ricerca Scientifica 1, 00133, Rome (Italy) \\
  \texttt{fabio.bonaccorso@roma2.infn.it} \\
   \and
  Niccolò Cocciaglia \\
  Department of Physics \& INFN, \\ 
  Tor Vergata University of Rome, \\ 
  Via della Ricerca Scientifica 1, 00133, Rome (Italy) \\
  \texttt{niccolo.cocciaglia@roma2.infn.it} \\
   \and
  Robin A. Heinonen \\
  MaLGa \& DICCA\\
  University of Genoa, \\
  Via Montallegro 1, Genoa (Italy)\\
  \texttt{robin.alrik.heinonen@edu.unige.it} 
   \and 
  Lorenzo Piro \\
  Department of Physics \& INFN, \\ 
  Tor Vergata University of Rome, \\ 
  Via della Ricerca Scientifica 1, 00133, Rome (Italy) \\
  \texttt{lorenzo.piro@roma2.infn.it} 
}

\begin{document}
\maketitle

\begin{abstract}

Identifying the location and characteristics of pollution sources in turbulent flows is challenging, especially for environmental monitoring and emergency response, due to sparse, stochastic, and infrequent cue detection. 
Even in idealized settings, accurately modeling these phenomena remains highly complex, with realistic representations typically achievable only through experimental or simulation-based data.

We introduce TURB-Smoke, a cutting-edge numerical dataset designed for investigating odor and contaminant dispersion in turbulent environments with and without mean wind. 
Generated via direct numerical simulations of the fully resolved three-dimensional Navier–Stokes equations, TURB‑Smoke tracks hundreds of millions of Lagrangian particles released from five distinct point sources in fully developed turbulence, thus providing a reliable ground-truth framework for developing and evaluating source-tracking strategies using stationary sensors or mobile agents in realistic flows.

Each particle's trajectory is continuously tracked on many characteristic turbulence timescales, recording both the position and the local flow velocity. Additionally, we provide coarse-grained concentration fields in 3D and in quasi-2D slabs containing the source, ideal for quickly testing and optimizing search algorithms under varying flow conditions.
 
\end{abstract}

\section{Background \& Summary}
Atmospheric~\cite{fernando2010,gonzalez2021} and aquatic~\cite{nanoplastics, oil_plastics} pollutants pose grave threats to ecosystems and human health~\cite{threat_nanoplastics, lelieveld2015premature}. 
Predicting the dynamics of pollutants, containing their spread, and locating their sources are difficult and critical tasks that are extremely sensitive to the physics governing their mixing into the background flow.  In both oceanic and atmospheric contexts, this flow is generically turbulent, with profound and complex consequences for the dispersion of pollutants and other chemicals such as odors~\cite{shraiman2000}.
Physically, chemicals immersed in a fluid can be modeled as passive scalars~\cite{warhaft} and are transported by a combination of flow advection and molecular diffusion. 
Chemicals organize into a statistically steady but constantly fluctuating landscape~\cite{celani2014} consisting of a correlated, evolving ensemble of puffs and filaments. 
Such landscapes exhibit strong spatiotemporal intermittency and nontrivial, non-Gaussian statistics such as power-law distributions of the timings between consecutive large-concentration events. Even in idealized scenarios, the precise mathematical modeling of these phenomena is extremely challenging, and realistic descriptions are currently only accessible through data from experiments and simulations.

Beyond its relevance to environmental science, the dispersion of chemicals in turbulent flows also enjoys important applications in biology. 
Numerous animals have evolved strategies to track sources of chemical odors in geophysical flows, in the service of mating, foraging, or predation --- a problem usually referred to as ``olfactory search''~\cite{reddy}. 
These include aquatic animals such as plankton~\cite{roozen2001} and lobsters~\cite{derby2014}, flying insects such as moths and fruit flies~\cite{murlis1992} (which instead search in an atmospheric setting), and mammals such as mice~\cite{schultz1973} and dogs~\cite{dogs} that integrate odor signals from the ground and air during search~\cite{rigolli2022}. 
In all these settings, the properties and physics of the large-scale structure formed by the advected odor, called a plume, have crucial effects on the efficacy of various search strategies and, therefore, are intimately connected to the behavior of animals~\cite{balkovsky2002,durve2020}. 

Motivated and inspired by ethology, a wide range of heuristic search strategies have been developed. 
A paradigmatic example is infotaxis~\cite{vergassola2007}, a strategy framed in the context of Bayesian inference that has shown robust performance in a variety of environments for both single and multi-agent search~\cite{heinonen2023,panizon2023,piro2025-arxiv}. 
These approaches typically operate under the assumption that the ``agents'' can use a model of the environment to interpret their encounters with the contaminant, which are assumed to be statistically independent, thus neglecting the strong spatiotemporal correlations inherent to turbulent transport that can substantially alter search efficiency~\cite{heinonen2025exploring}.
To bypass the need for such assumptions, recent data-driven methods based on reinforcement learning (RL) have been developed, enabling individual agents to learn effective search policies directly from experience in complex and turbulent  landscapes~\cite{singh2023,rando2025}. 

Thus, access to high-quality data is crucial for the serious study of the various problems associated with chemical dispersion in turbulent flows. 
To our knowledge, no such data are currently available to the public. Therefore, in this work, we present TURB-Smoke~\cite{turbsmoke}, a database generated from high-fidelity direct numerical simulation (DNS) of a turbulent flow with fixed sources of (passive) pollutants. 
The chemicals are modeled as Lagrangian tracer particles (neglecting the relatively unimportant effect of molecular diffusion), whose positions and velocities are tracked in time. 
Both raw Lagrangian data and coarse-grained Eulerian concentration fields are provided in the database. 
To model oceanic currents and atmospheric wind, we include data with and without a background mean flow with several different magnitudes. In addition, each flow contains five independent particle sources, allowing the study of problems involving mixing of multiple contaminants.
These data have been used in several publications devoted to developing and testing olfactory search strategies~\cite{heinonen2023,heinonen2025exploring,piro2025-jot,piro2025-arxiv}, but the (physical) framework built for such investigations also offers a unique idealized test bed for many realistic atmospheric or oceanic environments contaminated by polluting substances or hazardous gas leaks. 
Our database has great potential utility for environmental applications, one of many being the design of robot strategies to detect and locate source fluid leakages~\cite{kong2019,tariq2021}. 



\section{Methods}


All the data presented here are computer generated, so no experimental procedure is involved. 
The background turbulent flow is reproduced in a cubic, periodic domain with a pseudospectral simulation of the Navier-Stokes equation. 
The flow is seeded with particles, emitted in ``puffs'' that are passively advected by fluid motion and are further displaced by a constant mean wind. 
The 3D and 2D concentration fields are then obtained by counting particles on coarser grids. 
The following subsections present more detailed descriptions of the methods, along with all numerical parameters.


\subsection{Simulation of the turbulent flow}
\label{subsec:methods_DNS}
\begin{figure}[t]
    \centering
    \begin{minipage}[t]{0.45\textwidth}
        \centering
        \includegraphics[width=\linewidth]{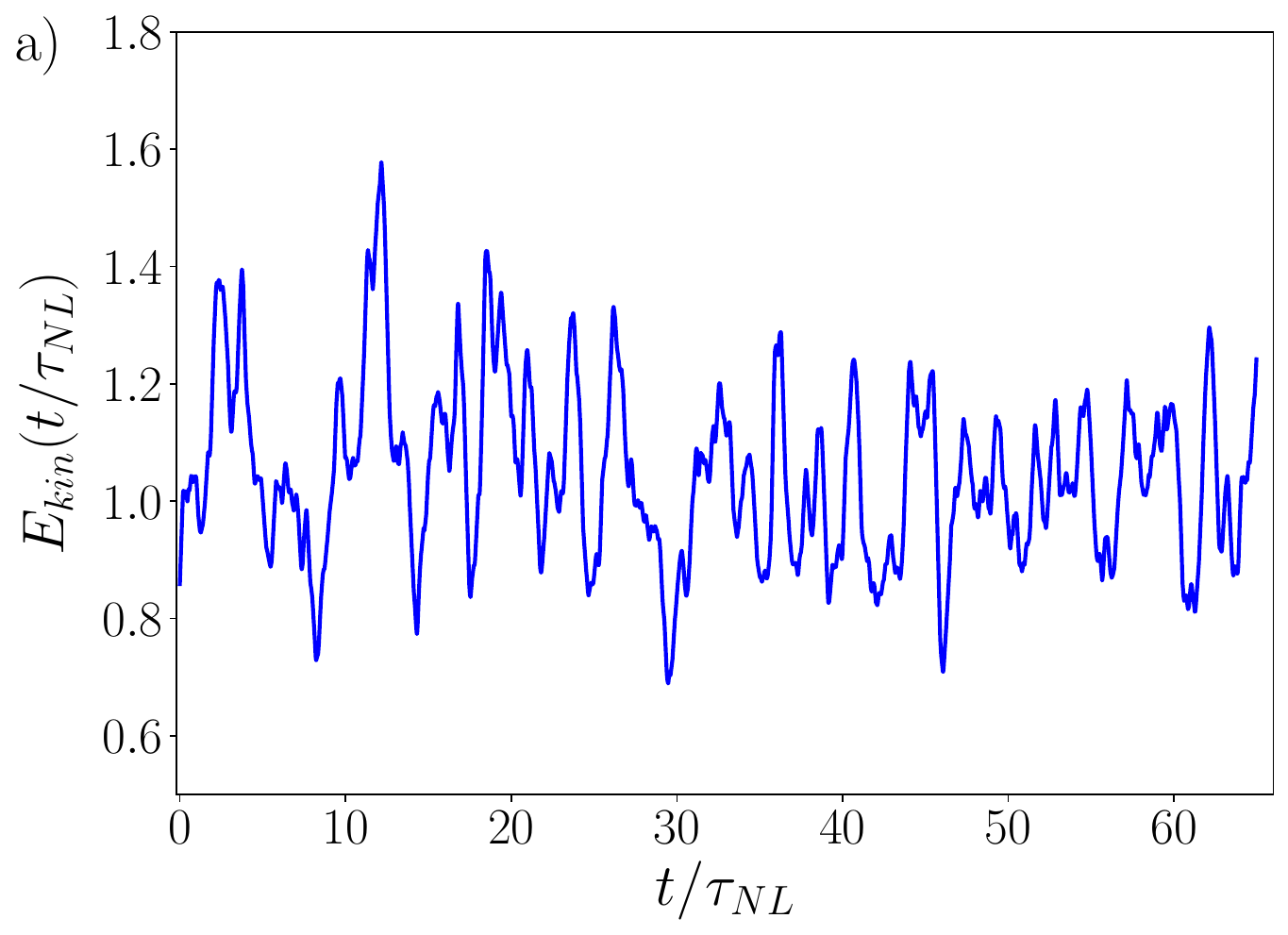}
        \hfill
        \includegraphics[width=\linewidth]{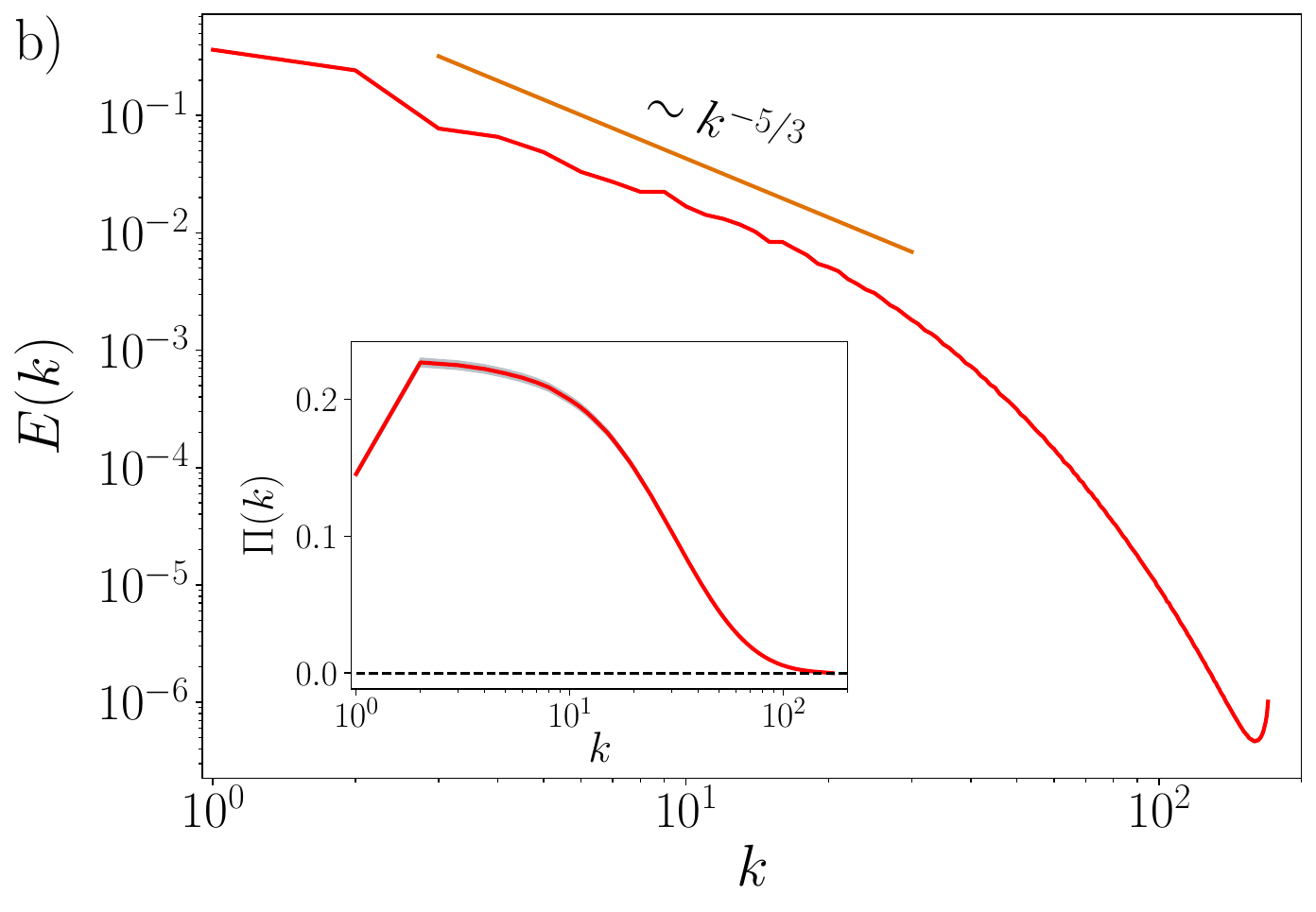}
    \end{minipage}
    \hfill
    \begin{minipage}[t]{0.53\textwidth}
        \centering
        \raisebox{-2.8cm}{\includegraphics[width=\linewidth]{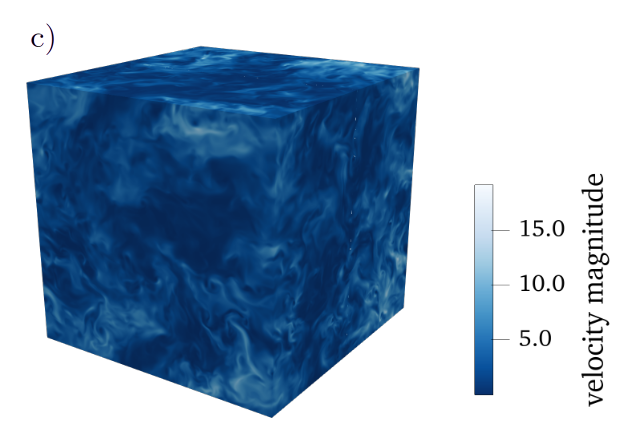}}
    \end{minipage}
\caption{(a) Time evolution of the mean kinetic energy $E_{kin}$ obtained from the RUN1 DNS, displaying the statistical stationarity of the signal. Time is measured in terms of the large-eddy (non-linear) turnover time $\tau_{NL}$.
(b) Kinetic energy spectrum $E(k)$ in the main panel and kinetic energy flux $\Pi(k)$ in the inset, with errors (grey shaded area). The instantaneous values of both spectrum and flux from the numerical simulation have been averaged using a block-averaging technique, namely splitting the data in $\sim 60$ blocks and computing mean values and errors, for each $k$, from the corresponding block means. In the main panel the errors are smaller than the red line.  
(c) Snapshot of the cubic domain with color coding representing the magnitude of the velocity field.}
\label{fig:enekin_and_stuff}
\end{figure}

The first step of our procedure is to reproduce a turbulent environment into which the Lagrangian particles (i.e. particles transported by the underlying flow) will then be inserted.
To this end, we performed a direct numerical simulation (DNS) of the incompressible Navier-Stokes (NS) equations. 
The NS equations read:
\begin{equation}
\label{eq:nse}
\begin{cases}
\partial_t \bm{u} + (\bm{u} \cdot \nabla) \bm{u} = - \bm{\nabla} p/\rho_0  
+ \nu\Delta\bm{u} + \mathbf{F} \\
\bm{\nabla} \cdot \bm{u} = 0,
\end{cases}
\end{equation}
where $\bm{u}=\bm{u}(\bm{r},t)$ is the velocity field, $p$ is the pressure, $\rho_0$ the mean density of the fluid and $\nu$ its kinematic viscosity. 
$\mathbf F$ is the homogeneous and isotropic forcing that drives the system to a non-equilibrium steady state. 
The divergence-free condition in \eqref{eq:nse} enforces incompressibility.
The domain is a cube of size $L=2\pi$, with periodic boundary conditions. 
The code follows a standard pseudospectral approach~\cite{ferzigerCFD}, in which multiplications are performed in a real (or physical) space while derivatives are computed in Fourier space. 
To prevent aliasing errors from higher wave numbers - a common problem when dealing with Fourier transforms - a dealiasing procedure with the $2/3$ rule is used. 
Time integration has been implemented with a second-order Adams-Bashforth scheme. 
The cubic grid that discretizes the physical space is made up of $N^3=512^3$ collocation points. 
A statistically steady turbulent state is maintained by forcing at large scales, specifically at wave numbers $0.5\leq k_{f}\leq 1.5$, via a second-order Ornstein-Uhlenbeck process~\cite{forcingsawford}.  
The correlation time of the forcing is comparable to the Kolmogorov time scale $\tau_\eta$, associated with the small turbulent scale (Kolmogorov scale $\eta$) where energy dissipation dominates. 
To quantify how much the flow is turbulent, we estimate the Reynolds number on the Taylor scale $Re_\lambda=U_{\text{rms}}\, \lambda/\nu$~\cite{frisch1995turbulence} where $U_{\text{rms}}$ is the root mean square value of the velocity field, and $\lambda = \sqrt{5E_{kin}/\Omega}$ is the Taylor scale measured in terms of the mean kinetic energy $E_{kin} = 1/V \int_V d\bm{r}\, |\bm{u}(\bm{r})|^2/2$ and the mean enstrophy $\Omega = 1/V \int_V d\bm{r}\, |\bm{\omega}(\bm{r})|^2/2$, where $\bm{\omega}(\bm{r}) = \bm{\nabla} \times \bm{u}(\bm{r})$ is the vorticity field. 
For reference, in state-of-the-art numerical simulations $Re_\lambda$ reaches values of $O(10^3)$, while in atmospheric turbulence it is estimated to achieve $O(10^4)$.
\newline 
For computational efficiency - especially when particles are also taken into account - two separate but statistically equivalent runs, referred to as RUN1 and RUN2, have been produced. 
In Table~\ref{tab:parameters} we report all the parameters and useful quantities that characterize our simulations. 
The averaged quantities reported there are based on both runs. 
In Figure~\ref{fig:enekin_and_stuff}(a) we show the time evolution of the stationary mean kinetic energy (per unit mass) $E_{kin}(t)$, for RUN1. 
Figure~\ref{fig:enekin_and_stuff}(b) displays, in the main panel, the (isotropic) kinetic energy spectrum $E(k)$, averaged in time and over a spherical shell in Fourier space of width $\Delta k = 2\pi /L$. 
It represents the kinetic energy density stored at scale $\sim 1/k$ and is defined as~\cite{alexakisbiferale}: 
\begin{equation}
    E(k) = (2\, \Delta k)^{-1} \sum_{k \leq \bm{k} \leq k+\Delta k} \langle|\tilde{\bm{u}}(\bm{k})|^2 \rangle_t
\end{equation}
where $\tilde{\bm{u}}(\bm{k})$ is the Fourier mode of the velocity field $\bm{u}(\bm{r})$ and $\langle\cdot\rangle_t$ denotes time averaging.
The inset shows the kinetic energy flux across spherical shells of radii $k$, which reads~\cite{alexakisbiferale}:
\begin{equation}
    \Pi(k) = \langle \bm{u}^{<k} \cdot [(\bm{u} \cdot \bm{\nabla})\bm{u}]\rangle_{\bm{r},t}.
\end{equation} 
Here $\bm{u}^{<k}(\bm{r}) = \sum_{|\bm{k}|\leq k} \tilde{\bm{u}}(\bm{k})\, e^{i \bm{k}\cdot\bm{r}}$ is the low-pass-filtered velocity field while $\langle\cdot\rangle_{\bm{r},t}$ indicates a spatio-temporal average.
Some characteristic features of 3D turbulence are observed. 
In the inertial range of scales, namely far enough from the forcing- and dissipation-dominated ones, the spectrum is close to the Kolmogorov scaling $E(k) \sim k^{-5/3}$. 
The energy flux is positive, indicating a direct (i.e. toward small scales) kinetic energy cascade. 
The inertial range is rather limited in our case: it increases as $Re_\lambda$ grows, and the latter is rather modest in our simulation. 
Finally, Fig.~\ref{fig:enekin_and_stuff}(c) is a snapshot taken from RUN1 that represents the magnitude of the velocity in the whole domain. Energetic thin structures, excited by external forcing, can be recognized on the domain surface.

\begin{figure}[t]
    \centering
    \includegraphics[width=\linewidth]{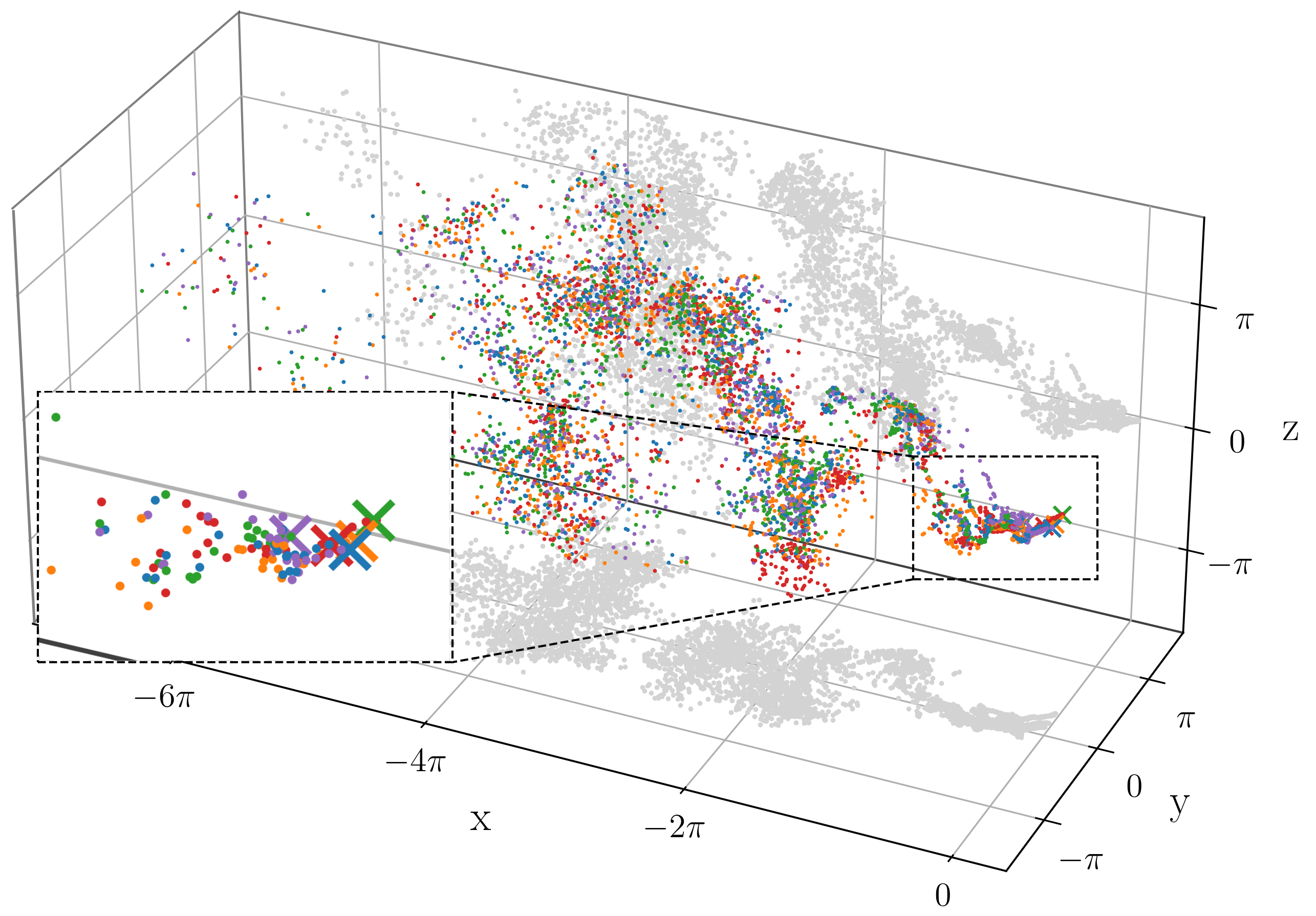}
    \caption{  Snapshot of the Lagrangian particles advected by the mean wind of intensity $U_0/U_\text{rms} = 2.05$. The snapshot is taken $9000$ time steps after the emission of the first ``puff'', or when $900$ puffs have been already emitted. The particles are colored according to their sources, that are depicted as colored crosses. A zoom on the emitting region is added, with less particles for better visibility. $xy$ and $xz$ projections (shadows) of the smoke are depicted in grey. Source $1$ (blue cross) has coordinates $\bm{s}_1=(0,0,0)$.}
    \label{fig:Lagr_wind_snapshot}
\end{figure}

\begin{table}[h!]
\centering
\begin{tabular}{|ccccc|}
\hline

$N$  &  $L$  &  $dx$  &  $dt$  &  $\nu$ \\
$512$ & $2\pi$  &  $\simeq 1.23\times 10^{-2}$  &  $5.00\times 10^{-4}$  &  $1.25 \times 10^{-3}$  \\ \hline
$U_\text{rms}$  &  $\tau_{NL}$  &  $\epsilon$  &  $\tau_\eta$  &  $Re_\lambda$  \\
$1.461 \pm 0.008$  &  $4.30 \pm 0.02$  &  $0.233 \pm 0.004$  &  $(74.5 \pm 0.7)\times 10^{-3}$  &  $166 \pm 3$  \\ \hline

\end{tabular}
\caption{\color{black}Parameters of the DNS. $N$ is the number of grid points in each dimension, $L$ the physical dimension of the periodic box, $dx=L/N$ the grid spacing, $dt$ the time integration step, $\nu$ the kinematic viscosity, $U_\text{rms}$ is the root-mean-square turbulent velocity, $\tau_{NL}=U_{rms}/L_f$, with $L_f = 2\pi/k_f = 2\pi$, is the large-eddy (or nonlinear) turnover time, $\epsilon = \nu \langle \partial_i u_j \partial_i u_j \rangle$ the (space-time) average energy dissipation rate, 
$\tau_\eta = \sqrt{\nu/\epsilon}$ the Kolmogorov timescale and $Re_\lambda$ the Taylor-scale Reynolds number. The mean values and errors of the fluctuating quantities are computed from a block-averaging procedure on both RUN1 and RUN2 (see caption of Figure~\ref{fig:enekin_and_stuff}).
}
\label{tab:parameters}
\end{table}


\subsection{Lagrangian particles}
\label{subsec:methods_Lagr}

The simulations described in the previous subsection produced a Homogeneous Isotropic Turbulent (HIT) flow. 
However, we want our particles to feel the effect of a constant mean wind superimposed on the turbulent motion.
This is accomplished by imposing the wind directly on the particles' equations of motion. 
The HIT flow is thus seeded with noninteracting tracer particles that are passive, namely, have no back-reaction on the flow velocity. They are emitted by five different point sources and transported by the combined effect of the NS flow and the mean wind. 
Their position $\bm{X}(t)$ evolves according to the Lagrangian equation:
\begin{equation}
\begin{cases}
    d {\mathbf X}(t) / dt=\bm u(\mathbf X(t),t) + \boldsymbol{U}_0 \\
    \mathbf X(0) = \mathbf s_i
\end{cases}
    \label{eq:lagr_eq}
\end{equation}
where $\bm{U}_0 = -U_0\, \hat{x}$ is the wind velocity and $\mathbf s_i, i=1,\dots,5,$ denotes the position of the $i$-th particle source. 
Eqs.~\eqref{eq:lagr_eq} are integrated using a trilinear or B-spline 6th order interpolation scheme~\cite{hinsberg2012} to obtain the fluid velocity $\bm u$ at the particle position $\bm{X}(t)$.  
Particles are free to cross the periodic boundaries many times, but the turbulent flow they experience is always the one in the cube --- otherwise stated, a particle positioned at $\bm{X}(t)$ sees the same HIT field as a particle at $\bm{X}(t) + (iL,jL,kL)$, $(i,j,k) \in \mathbb{Z}^3$. 
However, particles are removed from the numerical domain after they have traveled a distance $10 L$ from their emission point. 

The relative positions of the sources and the parameters that describe the emission process are found in Table~\ref{tab:sources}.
The position of the first source $\mathbf{s}_1$ is arbitrary due to statistical homogeneity. The sources $\mathbf{s}_2$ and $\mathbf{s}_3$ are placed along the direction $\hat{y}$, perpendicular to the wind, while $\mathbf{s}_4$ and $\mathbf{s}_5$ are along the wind direction $\hat{x}$. 
The five sources are thus coplanar and lie in a plane perpendicular to the $z$ axis.
The particles are not generated exactly at the source positions, but are uniformly random inside spheres of radius $r=3\, dx$ centered at $\bm{s}_i$.
RUN1 refers to simulations with mean wind $U_0/U_\text{rms}=\{0,1.03\}$ while in RUN2 we have the stronger winds with $U_0/U_\text{rms}=\{2.05,4.11,6.16\}$. 
In Figure~\ref{fig:Lagr_wind_snapshot} a snapshot of the ``smoke" made of Lagrangian particles is represented, with different colors according to which source they are emitted by.

\begin{table}[h!]
\centering
\begin{tabular}{|c|c|c|}
\hline 
Quantity & Symbol & Value \\
\hline
Relative position of source 2 w.r.t. source 1 & $\mathbf{s}_2 - \mathbf{s}_1$ & $0.029L\, \hat{y}$ \\
Relative position of source 3 w.r.t. source 1 & $\mathbf{s}_3 - \mathbf{s}_1$ & $0.098L\, \hat{y}$ \\
Relative position of source 4 w.r.t. source 1 & $\mathbf{s}_4 - \mathbf{s}_1$ & $-0.098L\, \hat{x}$ \\
Relative position of source 5 w.r.t. source 1 & $\mathbf{s}_5 - \mathbf{s}_1$ & $-0.029L\, \hat{x}$ \\ 
Total number of particles emitted at every puff & $N_p$ & 1000 \\ 
Time delay between puffs & $t_p$ & $10\, dt$ \\
\hline
\end{tabular}
\vspace{3pt}
\caption{Parameters describing the sources and the emission process. Relative positions are measured in terms of the domain size $L$. The $N_p$ particles emitted by a single puff are equidistributed among the five sources (i.e. $N_p/5$ particles emitted by every source).}
\label{tab:sources}
\end{table}


\subsection{Concentration field in 3D}
\label{sec:3D_concentrations}
At this point, the data on the Lagrangian trajectories are used to obtain the 3D concentration field. 
The case with wind intensity $U_0/U_\text{rms}=2.05$ has been used to study olfactory search algorithms by some of us in recent studies~\cite{piro2025-jot, piro2025-arxiv}, and is the one included in the database.
A three-dimensional grid of $N_x\times N_y\times N_z=129\times99\times99$ cells is defined, with grid spacing $\Delta \simeq 12.2\, dx$. 
We may denote the concentration field as $c_{(i,j,k)}(t)$, in which $(i,j,k)$, $i=1,\dots,N_x$, $j=1,\dots,N_y$, $k=1,\dots,N_z$ indicates the position of a 3D grid element. 
The concentration at $(i,j,k)$ is calculated, at each time, as the number of particles whose $(x,y,z)$ coordinates fall into that cell.

\subsection{Concentration field in 2D}
\label{sec:2D_concentrations}
\begin{figure}[t]
    \centering
    \includegraphics[width=.9\linewidth]{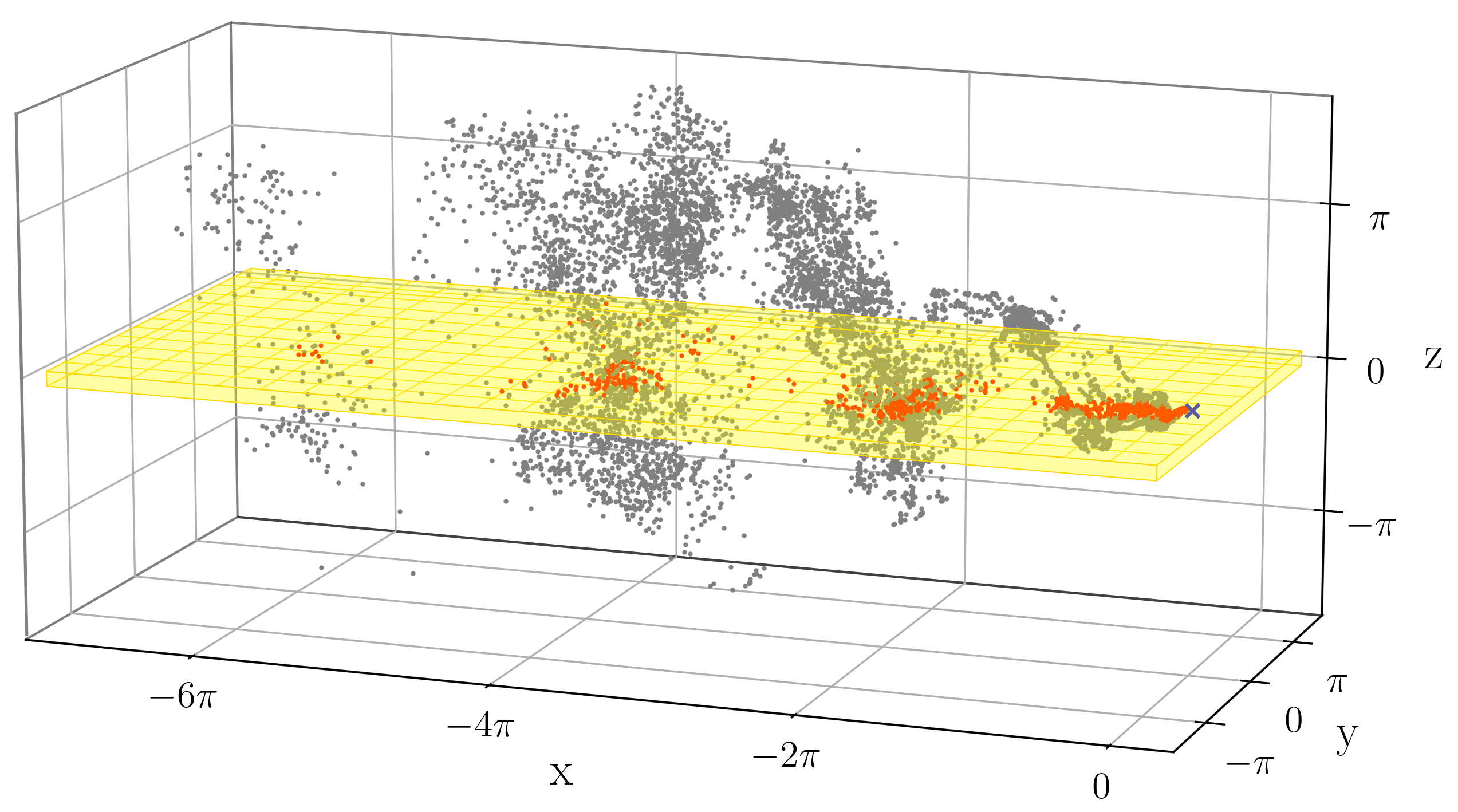}
    \caption{Representation of the filtering performed by the thin slab, used to obtain the 2D concentrations. Only the particles highlighted in red - those falling into the slab - are sampled within the 2D grid. The snapshot here displayed is the same of Figure~\ref{fig:Lagr_wind_snapshot}, but only particles emitted by the source in $\bm{s}_1$ (blue cross) are represented. Slab thickness and cells widths have been increased for clarity.}
    \label{fig:Lagr_wind_2D_plane}
\end{figure}

To obtain the values of the 2D concentrations, an analogous approach is followed, but now with a rectangular grid oriented perpendicularly to the $z$ axis, having resolution $N_x \times N_y$ points and the same grid spacing $\Delta$ of the 3D mesh. 
This 2D slice of the domain should be intended as a 3D grid with $N_z=1$, in the sense that a particle contributes to the concentration if $|Z(t)-s_{i,z}| < \Delta / 2$, where $Z(t)$ is the $z$ coordinate of a particle at time $t$ and $s_{i,z}$ the $z$ coordinate of source $i$. 
This thin slab contains the plane where the sources are located. 
As a visual reference, Figure~\ref{fig:Lagr_wind_2D_plane} shows the slab with the embedded 2D grid, selecting only some of the particles (red) among all emitted particles (gray).

\begin{figure}[t]
    \centering
    \includegraphics[width=\linewidth]{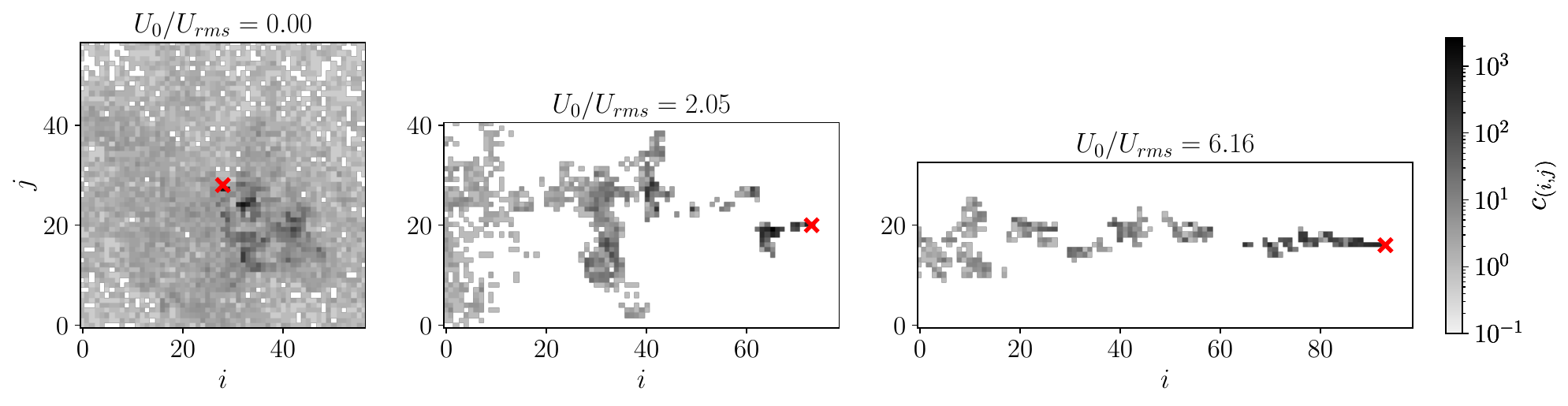}
    \caption{Three snapshots of the 2D concentration field $c_{(i,j)}$ generated from the source at $\bm{s}_0$ (red cross). Left, center and right panels refer to three values of the mean wind: zero, intermediate ($U_0/U_{rms}=2.05$) and large ($U_0/U_{rms}=6.16$). A logarithmic colorbar is used for more visual clarity.}
    \label{fig:2D_snapshots}
\end{figure}

Since all wind intensities are now considered, $N_x$ and $N_y$ vary with $U_0/U_\text{rms}$ as reported in Table~\ref{tab:2Dgrids}. 
\begin{table}[h!]
\centering
\begin{tabular}{|c|c|c|}
\hline 
$U_0/U_\text{rms}$  & $N_x$  & $N_y$ \\
\hline
0 & 57 & 57 \\ 
1.03 & 67 & 49 \\ 
2.05 & 79 & 41 \\ 
4.11 & 89 & 37 \\ 
6.16 & 99 & 33 \\ 
\hline
\end{tabular}
\vspace{3pt}
\caption{\color{black}Sizes of the 2D grid at varying intensity of the mean wind.}
\label{tab:2Dgrids}
\end{table}
The number of grid points $N_x \cdot N_y$ has been kept roughly constant at varying $U_0$ and, as the wind grows, the aspect ratio $N_x/N_y$ of the grid also increases. 
Here, the concentration $c_{(i,j)}(t)$ is calculated as the number of particles whose $(x,y)$ coordinates fall within that grid element. 
Figure~\ref{fig:2D_snapshots} shows three snapshots of the concentration field for zero, intermediate and large values of the mean wind.
 

\section{Data Records}


All data sets are hosted on the SMART-Turb portal (\href{http://smart-turb.roma2.infn.it}{http://smart-turb.roma2.infn.it}), and located in the TURB-Smoke database~\cite{turbsmoke}. 
Other datasets concerning rotating turbulence (TURB-Rot~\cite{TURB-Rot}), Lagrangian trajectories in HIT (TURB-Lagr~\cite{TURB-Lagr}), helically-forced turbulence (TURB-Hel~\cite{TURB-Hel}) and magnetohydrodynamic turbulence (TURB-MHD~\cite{TURB-MHD}) are already available on the website.

In the following, we provide further details about the specifics and describe the files inside the directories related to the three groups of data (Lagrangian, 3D, 2D). 

\begin{itemize}[leftmargin=4mm]
    \item[] \textbf{Lagrangian data -} 
    Raw data of the generated Lagrangian particles for three mean-wind intensities, $U_0/U_\text{rms}=\{0,\,2.05,\,6.16\}$, specifically their ID, positions and velocities, have been stored every $\Delta t=150\, dt \simeq \tau_\eta$. 
    Here and in the following the three cases will be also referred to as zero-wind, intermediate-wind and strong-wind, respectively.
    At variance with the concentrations fields, velocities and positions are measured in the reference frame of an ``external observer'', who sees no mean wind but rather the particle sources being displaced with constant velocity $-\boldsymbol{U}_0=U_0 \hat{x}$, in the same way a person on a train platform would see the smoke emitted from a moving steam train. 
    Otherwise stated, the data describe the motion of particles originating from sources moving at constant velocity and advected by HIT. 
    A simple Galilean transformation, performed e.g. for producing Figures~\ref{fig:Lagr_wind_snapshot} and \ref{fig:Lagr_wind_2D_plane}, allows to switch to the reference frame ``comoving'' with the sources.
    The particles move in the periodic domain until they have traveled a distance $10L$, after which they are removed. 
    The Lagrangian datasets with zero and non-zero mean velocities are in the \verb+Lagrangian_data+ directory, in the subdirectories \verb+data_wind0+, \verb+data_wind205+ and \verb+data_wind616+ subdirectories, respectively. 
    The numbers in the directory names, also used for data files, correspond to the quantity $100\,U_0/U_\text{rms}$.
    Both folders contain .h5 files named ``Lagr\_\textit{x}wind\_\textit{y}.h5'', where \textit{y} is a multiple of $150$ and \textit{x}=$\{0,205,616\}$, storing particles data at timestep \textit{x}. 
    Each of these files describes particles emitted by all five sources: an explanation on how to associate particles to their own sources is found in the text file ``discriminating\_sources\_and\_puffs.txt''.
    For all cases we collected $1501$ snapshots covering a time period $T\simeq 26\, \tau_{NL}$. 
    It is worth pointing out that, in the absence of mean wind, the particles need a very long time to reach the cut-off distance $10\, L$ from their sources. 
    As a consequence the number of particles keeps on increasing with time, and in the latest file there are approximately $22.5$ millions of particles stored. 
    
    \item[] \textbf{3D concentrations -}
    The 3D concentration field on the coarser grid $N_x \times N_y \times N_z$ has been stored every $\Delta t$ and for a total time $T \simeq 53\, \tau_{NL}$, resulting in approximately $3000$ snapshots.  
    The 3D concentrations dataset is provided in the \verb+3D_concentrations+ directory. 
    Here five .h5 files can be found, named ``3D\_conc\_source\textit{x}'', where \textit{x}$\in[0,4]$ indicates one of the five particle sources. 
    Thus, at variance with the Lagrangian case, here the data of all snapshots are stored together in one file for each source.
    
    \item[] \textbf{2D concentrations -}
    Similarly, we computed the 2D concentrations on the $N_x \times N_y$ grids (as in Tab.~\ref{tab:2Dgrids}) with the same time stepping $\Delta t$, for a total time that varies with the wind intensity (low speeds have fewer time snapshots), ranging from $T \simeq 41\, \tau_{NL}$ to $T \simeq 53\, \tau_{NL}$. The number of successive configurations stored ranges between $\sim 2500$ to $\sim 3000$. 
    The dataset is provided in the \verb+2D_concentrations+ directory, in Python~.pkl format. 
    There are two files for each of the five wind intensities $U_0/U_\text{rms}=\{0,\, 1.03,\, 2.05,\, 4.11,\, 6.16\}$, named ``2D\_conc\_wind\textit{x}'' and ``2D\_conc\_wind\textit{x}\_interpolated'', with \textit{x}$=\{0,\, 103,\, 205,\, 411,\, 616\}$. The former set of files contains one snapshot every $\Delta t$ and was used in a recent study~\cite{heinonen2025lowinfo}. The latter, used in another publication~\cite{heinonen2025exploring}, contains one snapshot for every $\Delta t/10$ in order to access finer timescales than the DNS dumping frequency. This was achieved by linearly interpolating the Lagrangian position data before coarse-graining. When interpolation was impossible (i.e., because the particle had just been emitted by the source), the position of a particle was extrapolated backward in time using its instantaneous velocity and acceleration, and destroyed when the extrapolation brought it within a small radius around the source. The two datasets have some small inconsistencies due to the interpolation and are therefore treated as independent.

\end{itemize}

\section{Technical Validation}

To demonstrate the quality of the Lagrangian dataset, from which the various concentrations are derived, we decide to focus on the intermittent statistics displayed by the time increments of the velocity components $\Delta_\tau V_i \equiv V_i(t+\tau) - V_i(t)$, where $V_i$ is the $i$-th velocity component of a given particle. 

The moments of the probability distribution of such quantities, or Lagrangian structure functions~\cite{Arn_odo_2008}
\begin{equation}
    S_i^{(p)}(\tau) = \langle (\Delta_\tau V_i)^p \rangle\ 
    \label{eq:lagr_sf}
\end{equation}
show that large statistical deviations occur for small time-separations, while Gaussian-like statistics are recovered at large enough $\tau$. 
Usually intermittency reveals itself in the
deviation from self-similar scaling of the structure functions, namely $S_i^{(p)}(\tau) \sim \tau^{\xi(p)}$ where the ``anomalous exponents ''$\xi(p)$ are non-linear functions of $p$. 
To avoid any fitting procedures and any hypothesis of power law behavior~\cite{biferale2008lagrangianSF}, and following the concept of ``extended'' self-similarity~\cite{benzi1993extended}, we focus on the logarithmic derivative of $S_i^{(p)}(\tau)$ with respect to $S_i^{(2)}(\tau)$, referred to as the local slope or local exponent:
\begin{equation}
    \zeta_i(p,\tau) = \frac{d \log S_i^{(p)}(\tau)}{d \log S_i^{(2)}(\tau)}\ .
    \label{eq:local_slope_i}
\end{equation}
\begin{figure}[t]
    \centering
    \includegraphics[width=0.8\linewidth]{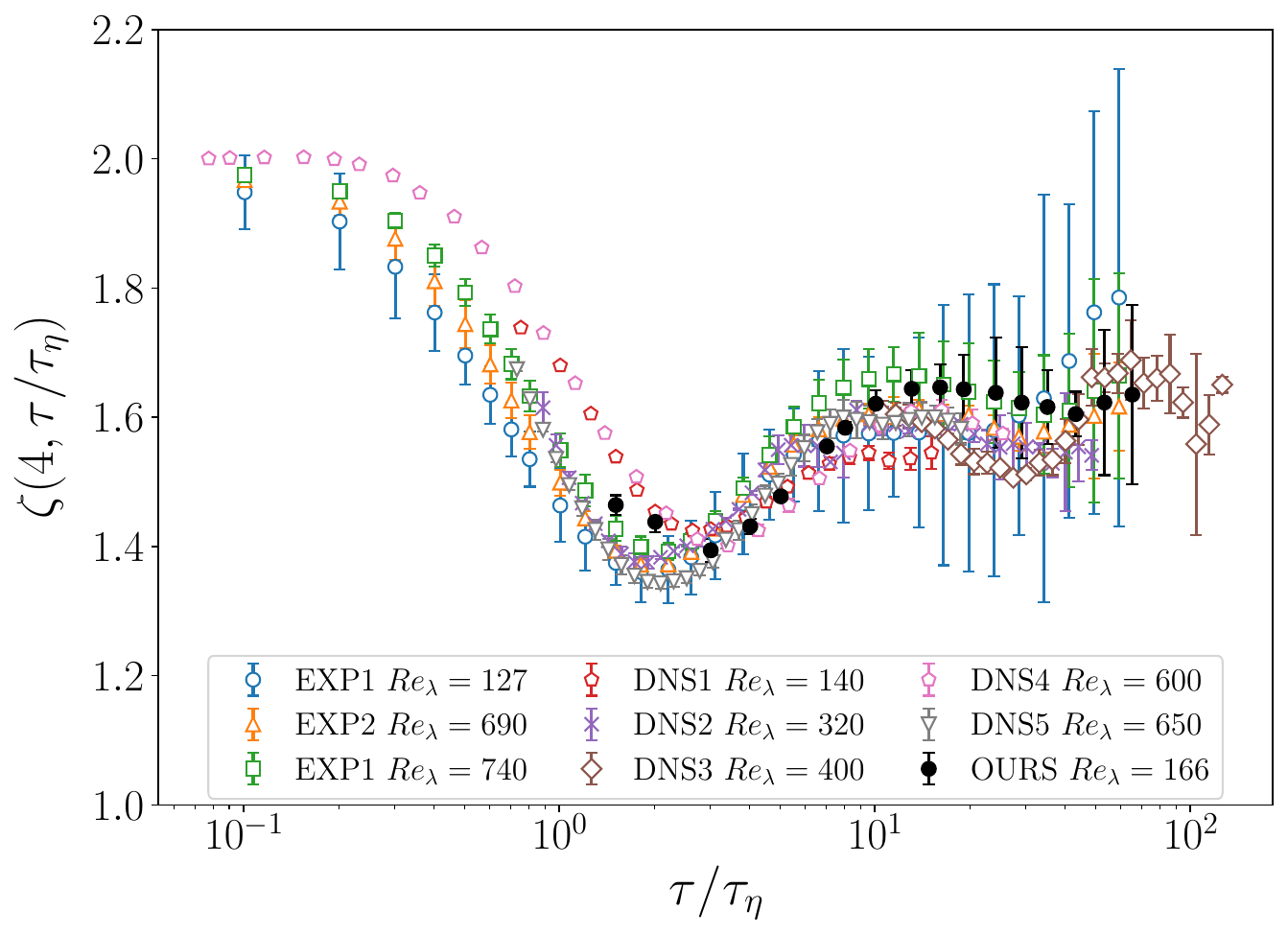}
    \caption{Fourth-order local slope $\zeta(4,\tau/\tau_\eta)$, computed as the component-wise average of the local slopes along $i$ (see Eq.~\eqref{eq:local_slope_i}). The result from our data (black filled circles) is compared with three experimental -- EXP1~\cite{EXP1}, EXP2~\cite{EXP2}, EXP3~\cite{EXP3} in the legend -- and five numerical -- DNS1~\cite{DNS1}, DNS2~\cite{DNS2}, DNS3~\cite{DNS3}, DNS4~\cite{DNS4}, DNS5~\cite{DNS5} in the legend -- measurements. Error bars in our data are computed as the absolute dispersions along the three directions. Adapted from Arnèodo et al.~\cite{Arn_odo_2008}.}
    \label{fig:local_slope}
\end{figure}
We report in Figure~\ref{fig:local_slope} the component-wise average for $p=4$ from our data (full black circles), compared to a collection of measurements of the same quantity from other state-of-the-art experimental and numerical studies~\cite{Arn_odo_2008}. 
The larger error bars for large $\tau/\tau_\eta$ can be attributed to higher large-scale anisotropy effects, due to the specific point source emission setup that we analyzed in this paper.
The agreement with the other data is evident, and the deviation of the local slope from its non-intermittent, self-similar value $\zeta(4,\tau) = 2$ is consistent for all time lags, but stronger for $\tau\sim\tau_\eta$, as expected from small-scale intermittency. 
We point out that the computation of the local slope requires the knowledge of $\log S_i^{(p)}(\tau)$ for close-by lag values $\tau$. 
Since our data are saved every $t\simeq \tau_\eta$, the minimum time separation we can obtain for $\zeta$ is the midpoint of $\tau/\tau_\eta \simeq 1$ and $\tau/\tau_\eta \simeq 2$. 
For this reason, $\tau/\tau_\eta<1.5$ is not available from this specific dataset.

    \label{eq:flatness}

\section{Usage Notes}


All datafiles in the repository are saved in a hierarchical binary format, either in .h5 or .pkl type. 
In order to help users access, visualize and use our data, we added a \verb+support+ directory in the database, containing the Jupyter notebook scripts used to generate Figures~\ref{fig:Lagr_wind_snapshot}, \ref{fig:Lagr_wind_2D_plane} and \ref{fig:2D_snapshots}. 
They show how to read, in Python, datasets saved in .h5 or in .pkl format.
The notebooks can be identified by the names ``Fig\textit{x}\_*.ipynb'', with \textit{x} being the figure number in this documentation and * a wildcard string. 
In the same folder, the interested user will find instructions to read 3D concentration fields in the notebook ``Plot\_3D\_concentrations.ipynb''.

\section{Code Availability}
All data presented in this paper are obtained by computer simulations using software developed in-house by the research group. We provide an executable for the Ubuntu/Linux operating system that can replicate all the results, with suitable computing power and time. The executable is released under the 3-Clause BSD License~\cite{3bd-clause}.

The data analysis code to generate the plots is in Python language and is provided, under the 3-Clause BSD License, on the SMART-Turb portal (\href{http://smart-turb.roma2.infn.it}{http://smart-turb.roma2.infn.it}) in the \verb+support+ folder of the TURB-Smoke dataset~\cite{turbsmoke}.

\section{Author Contributions}
LB, FB, RAH, and LP conceived the research and produced the data with numerical simulations. 
NC cleaned the data and verified their integrity. 
FB and NC wrote the scripts to read and visualize the data. 
NC, RAH, and LP wrote the manuscript. 
All authors reviewed the manuscript.

\section{Competing Interests}
The authors declare no competing interests.

\section{Acknowledgements}
We acknowledge useful discussions with Mauro Sbragaglia and  Massimo Vergassola. We also acknowledge financial support under the National Recovery and Resilience Plan (NRRP), Mission 4, Component 2, Investment 1.1, Call for tender No. 104 published on 2.2.2022 by the Italian Ministry of University and Research (MUR),
funded by the European Union – NextGenerationEU– Project Title ``Equations informed and data-driven approaches for collective optimal search in complex flows (CO-SEARCH)'', Contract 202249Z89M. – CUP B53D23003920006 and E53D23001610006. 
This work was supported by the European Research Council (ERC) under the European Union’s Horizon 2020 research and innovation program Smart-TURB (Grant Agreement No. 882340). RAH is supported by European Research Council under the grants Smart-TURB (No. 882340)
and RIDING (No. 101002724), by the Air Force Office of Scientific Research (grant FA8655-20-1-7028),
and the National Institute of Health under grant R01DC018789.


\bibliographystyle{naturemag}
\bibliography{references}

\begin{thebibliography}{10}
\expandafter\ifx\csname url\endcsname\relax
  \def\url#1{\texttt{#1}}\fi
\expandafter\ifx\csname urlprefix\endcsname\relax\def\urlprefix{URL }\fi
\providecommand{\bibinfo}[2]{#2}
\providecommand{\eprint}[2][]{\url{#2}}

\bibitem{fernando2010}
\bibinfo{author}{Fernando, H. J.~S.} \emph{et~al.}
\newblock \bibinfo{title}{Flow, turbulence, and pollutant dispersion in urban atmospheres}.
\newblock \emph{\bibinfo{journal}{Phys. Fluids}} \textbf{\bibinfo{volume}{22}}, \bibinfo{pages}{051301} (\bibinfo{year}{2010}).

\bibitem{gonzalez2021}
\bibinfo{author}{Gonz{\'a}lez-Mart{\'\i}n, J.}, \bibinfo{author}{Kraakman, N. J.~R.}, \bibinfo{author}{P{\'e}rez, C.}, \bibinfo{author}{Lebrero, R.} \& \bibinfo{author}{Mu{\~n}oz, R.}
\newblock \bibinfo{title}{A state--of--the-art review on indoor air pollution and strategies for indoor air pollution control}.
\newblock \emph{\bibinfo{journal}{Chemosphere}} \textbf{\bibinfo{volume}{262}}, \bibinfo{pages}{128376} (\bibinfo{year}{2021}).

\bibitem{nanoplastics}
\bibinfo{author}{ten Hietbrink, S.}, \bibinfo{author}{Materić, D.}, \bibinfo{author}{Holzinger, R.}, \bibinfo{author}{Groeskamp, S.} \& \bibinfo{author}{Niemann, H.}
\newblock \bibinfo{title}{Nanoplastic concentrations across the {N}orth {A}tlantic}.
\newblock \emph{\bibinfo{journal}{Nature}} \textbf{\bibinfo{volume}{643}}, \bibinfo{pages}{412–416} (\bibinfo{year}{2025}).

\bibitem{oil_plastics}
\bibinfo{author}{Yang, M.}, \bibinfo{author}{Zhang, B.}, \bibinfo{author}{Xin, X.}, \bibinfo{author}{Lee, K.} \& \bibinfo{author}{Chen, B.}
\newblock \bibinfo{title}{Microplastic and oil pollution in oceans: Interactions and environmental impacts}.
\newblock \emph{\bibinfo{journal}{Sci. Total Environ.}} \textbf{\bibinfo{volume}{838}}, \bibinfo{pages}{156142} (\bibinfo{year}{2022}).

\bibitem{threat_nanoplastics}
\bibinfo{author}{Liu, X.} \emph{et~al.}
\newblock \bibinfo{title}{The threats of micro- and nanoplastics to aquatic ecosystems and water health}.
\newblock \emph{\bibinfo{journal}{Nat. Water}} \textbf{\bibinfo{volume}{3}}, \bibinfo{pages}{764–781} (\bibinfo{year}{2025}).

\bibitem{lelieveld2015premature}
\bibinfo{author}{Lelieveld, J.}, \bibinfo{author}{Evans, J.}, \bibinfo{author}{Fnais, M.}, \bibinfo{author}{Giannadaki, D.} \& \bibinfo{author}{Pozzer, A.}
\newblock \bibinfo{title}{The contribution of outdoor air pollution sources to premature mortality on a global scale}.
\newblock \emph{\bibinfo{journal}{Nature}} \textbf{\bibinfo{volume}{525}}, \bibinfo{pages}{367–371} (\bibinfo{year}{2015}).

\bibitem{shraiman2000}
\bibinfo{author}{Shraiman, B.~I.} \& \bibinfo{author}{Siggia, E.~D.}
\newblock \bibinfo{title}{Scalar turbulence}.
\newblock \emph{\bibinfo{journal}{Nature}} \textbf{\bibinfo{volume}{405}}, \bibinfo{pages}{639--646} (\bibinfo{year}{2000}).

\bibitem{warhaft}
\bibinfo{author}{Warhaft, Z.}
\newblock \bibinfo{title}{Passive scalars in turbulent flows}.
\newblock \emph{\bibinfo{journal}{Annu. Rev. Fluid Mech.}} \textbf{\bibinfo{volume}{32}}, \bibinfo{pages}{203--240} (\bibinfo{year}{2000}).

\bibitem{celani2014}
\bibinfo{author}{Celani, A.}, \bibinfo{author}{Villermaux, E.} \& \bibinfo{author}{Vergassola, M.}
\newblock \bibinfo{title}{Odor landscapes in turbulent environments}.
\newblock \emph{\bibinfo{journal}{Phys. Rev. X}} \textbf{\bibinfo{volume}{4}}, \bibinfo{pages}{041015} (\bibinfo{year}{2014}).

\bibitem{reddy}
\bibinfo{author}{Reddy, G.}, \bibinfo{author}{Murthy, V.~N.} \& \bibinfo{author}{Vergassola, M.}
\newblock \bibinfo{title}{Olfactory sensing and navigation in turbulent environments}.
\newblock \emph{\bibinfo{journal}{Annu. Rev. Condens. Matter Phys.}} \textbf{\bibinfo{volume}{13}}, \bibinfo{pages}{191--213} (\bibinfo{year}{2022}).

\bibitem{roozen2001}
\bibinfo{author}{Roozen, F.} \& \bibinfo{author}{L{\"u}rling, M.}
\newblock \bibinfo{title}{Behavioural response of {D}aphnia to olfactory cues from food, competitors and predators}.
\newblock \emph{\bibinfo{journal}{J. Plankton Res.}} \textbf{\bibinfo{volume}{23}}, \bibinfo{pages}{797--808} (\bibinfo{year}{2001}).

\bibitem{derby2014}
\bibinfo{author}{Derby, C.~D.} \& \bibinfo{author}{Weissburg, M.~J.}
\newblock \bibinfo{title}{The chemical senses and chemosensory ecology of crustaceans}.
\newblock \emph{\bibinfo{journal}{NHC}} \textbf{\bibinfo{volume}{3}}, \bibinfo{pages}{263--92} (\bibinfo{year}{2014}).

\bibitem{murlis1992}
\bibinfo{author}{Murlis, J.}, \bibinfo{author}{Elkinton, J.~S.}, \bibinfo{author}{Carde, R.~T.} \emph{et~al.}
\newblock \bibinfo{title}{Odor plumes and how insects use them}.
\newblock \emph{\bibinfo{journal}{Annu. Rev. Entomol.}} \textbf{\bibinfo{volume}{37}}, \bibinfo{pages}{505--532} (\bibinfo{year}{1992}).

\bibitem{schultz1973}
\bibinfo{author}{Schultz, E.~F.} \& \bibinfo{author}{Tapp, J.~T.}
\newblock \bibinfo{title}{Olfactory control of behavior in rodents}.
\newblock \emph{\bibinfo{journal}{Psychol. Bull.}} \textbf{\bibinfo{volume}{79}}, \bibinfo{pages}{21} (\bibinfo{year}{1973}).

\bibitem{dogs}
\bibinfo{author}{Thesen, A.}, \bibinfo{author}{Steen, J.~B.} \& \bibinfo{author}{Døving, K.~B.}
\newblock \bibinfo{title}{Behaviour of dogs during olfactory tracking}.
\newblock \emph{\bibinfo{journal}{J. Exp. Biol.}} \textbf{\bibinfo{volume}{180}}, \bibinfo{pages}{247--251} (\bibinfo{year}{1993}).

\bibitem{rigolli2022}
\bibinfo{author}{Rigolli, N.}, \bibinfo{author}{Reddy, G.}, \bibinfo{author}{Seminara, A.} \& \bibinfo{author}{Vergassola, M.}
\newblock \bibinfo{title}{Alternation emerges as a multi-modal strategy for turbulent odor navigation}.
\newblock \emph{\bibinfo{journal}{eLife}} \textbf{\bibinfo{volume}{11}}, \bibinfo{pages}{e76989} (\bibinfo{year}{2022}).

\bibitem{balkovsky2002}
\bibinfo{author}{Balkovsky, E.} \& \bibinfo{author}{Shraiman, B.~I.}
\newblock \bibinfo{title}{Olfactory search at high {R}eynolds number}.
\newblock \emph{\bibinfo{journal}{PNAS}} \textbf{\bibinfo{volume}{99}}, \bibinfo{pages}{12589--12593} (\bibinfo{year}{2002}).

\bibitem{durve2020}
\bibinfo{author}{Durve, M.}, \bibinfo{author}{Piro, L.}, \bibinfo{author}{Cencini, M.}, \bibinfo{author}{Biferale, L.} \& \bibinfo{author}{Celani, A.}
\newblock \bibinfo{title}{Collective olfactory search in a turbulent environment}.
\newblock \emph{\bibinfo{journal}{Phys. Rev. E}} \textbf{\bibinfo{volume}{102}}, \bibinfo{pages}{012402} (\bibinfo{year}{2020}).

\bibitem{vergassola2007}
\bibinfo{author}{Vergassola, M.}, \bibinfo{author}{Villermaux, E.} \& \bibinfo{author}{Shraiman, B.~I.}
\newblock \bibinfo{title}{‘{I}nfotaxis’ as a strategy for searching without gradients}.
\newblock \emph{\bibinfo{journal}{Nature}} \textbf{\bibinfo{volume}{445}}, \bibinfo{pages}{406--409} (\bibinfo{year}{2007}).

\bibitem{heinonen2023}
\bibinfo{author}{Heinonen, R.~A.}, \bibinfo{author}{Biferale, L.}, \bibinfo{author}{Celani, A.} \& \bibinfo{author}{Vergassola, M.}
\newblock \bibinfo{title}{Optimal policies for {B}ayesian olfactory search in turbulent flows}.
\newblock \emph{\bibinfo{journal}{Phys. Rev. E}} \textbf{\bibinfo{volume}{107}}, \bibinfo{pages}{055105} (\bibinfo{year}{2023}).

\bibitem{panizon2023}
\bibinfo{author}{Panizon, E.} \& \bibinfo{author}{Celani, A.}
\newblock \bibinfo{title}{Seeking and sharing information in collective olfactory search}.
\newblock \emph{\bibinfo{journal}{Phys. Biol.}} \textbf{\bibinfo{volume}{20}}, \bibinfo{pages}{065001} (\bibinfo{year}{2023}).

\bibitem{piro2025-arxiv}
\bibinfo{author}{Piro, L.}, \bibinfo{author}{Heinonen, R.~A.}, \bibinfo{author}{Carbone, M.}, \bibinfo{author}{Biferale, L.} \& \bibinfo{author}{Cencini, M.}
\newblock \bibinfo{title}{Policy heterogeneity improves collective olfactory search in 3-{D} turbulence}.
\newblock \bibinfo{note}{Preprint at \url{https://arxiv.org/abs/2504.11291} (2025)}.

\bibitem{heinonen2025exploring}
\bibinfo{author}{Heinonen, R.~A.}, \bibinfo{author}{Biferale, L.}, \bibinfo{author}{Celani, A.} \& \bibinfo{author}{Vergassola, M.}
\newblock \bibinfo{title}{Exploring {B}ayesian olfactory search in realistic turbulent flows}.
\newblock \emph{\bibinfo{journal}{Phys. Rev. Fluids}} \textbf{\bibinfo{volume}{10}}, \bibinfo{pages}{064614} (\bibinfo{year}{2025}).

\bibitem{singh2023}
\bibinfo{author}{Singh, S.~H.}, \bibinfo{author}{van Breugel, F.}, \bibinfo{author}{Rao, R. P.~N.} \& \bibinfo{author}{Brunton, B.~W.}
\newblock \bibinfo{title}{Emergent behaviour and neural dynamics in artificial agents tracking odour plumes}.
\newblock \emph{\bibinfo{journal}{Nat. Mach. Intell.}} \textbf{\bibinfo{volume}{5}}, \bibinfo{pages}{58--70} (\bibinfo{year}{2023}).

\bibitem{rando2025}
\bibinfo{author}{Rando, M.}, \bibinfo{author}{James, M.}, \bibinfo{author}{Verri, A.}, \bibinfo{author}{Rosasco, L.} \& \bibinfo{author}{Seminara, A.}
\newblock \bibinfo{title}{Q-learning with temporal memory to navigate turbulence}.
\newblock \emph{\bibinfo{journal}{eLife}} \textbf{\bibinfo{volume}{13}}, \bibinfo{pages}{RP102906} (\bibinfo{year}{2025}).

\bibitem{turbsmoke}
\bibinfo{author}{Biferale, L.}, \bibinfo{author}{Bonaccorso, F.}, \bibinfo{author}{Cocciaglia, N.}, \bibinfo{author}{Heinonen, R.~A.} \& \bibinfo{author}{Piro, L.}
\newblock \bibinfo{title}{{TURB-S}moke}.
\newblock \bibinfo{note}{\textit{Smart-TURB} \url{https://smart-turb.roma2.infn.it/init/routes/\#/logging/view_dataset/9/tabmeta} (2025)}.

\bibitem{piro2025-jot}
\bibinfo{author}{Piro, L.}, \bibinfo{author}{Heinonen, R.~A.}, \bibinfo{author}{Cencini, M.} \& \bibinfo{author}{Biferale, L.}
\newblock \bibinfo{title}{Many wrong models approach to localise an odour source in turbulence with static sensors}.
\newblock \emph{\bibinfo{journal}{J. Turbul.}} \textbf{\bibinfo{volume}{26}}, \bibinfo{pages}{153--173} (\bibinfo{year}{2025}).

\bibitem{kong2019}
\bibinfo{author}{Kong, Y.} \emph{et~al.}
\newblock \bibinfo{title}{Locating hazardous chemical leakage source based on cooperative moving and fixing sensors}.
\newblock \emph{\bibinfo{journal}{Sensors}} \textbf{\bibinfo{volume}{19}}, \bibinfo{pages}{1092} (\bibinfo{year}{2019}).

\bibitem{tariq2021}
\bibinfo{author}{Tariq, S.}, \bibinfo{author}{Hu, Z.} \& \bibinfo{author}{Zayed, T.}
\newblock \bibinfo{title}{Micro-electromechanical systems-based technologies for leak detection and localization in water supply networks: {A} bibliometric and systematic review}.
\newblock \emph{\bibinfo{journal}{J. Clean. Prod.}} \textbf{\bibinfo{volume}{289}}, \bibinfo{pages}{125751} (\bibinfo{year}{2021}).

\bibitem{ferzigerCFD}
\bibinfo{author}{Ferziger, J.~H.}, \bibinfo{author}{Perić, M.} \& \bibinfo{author}{Street, R.~L.}
\newblock \emph{\bibinfo{title}{Computational Methods for Fluid Dynamics}} (\bibinfo{publisher}{Springer Nature}, \bibinfo{year}{2020}).

\bibitem{forcingsawford}
\bibinfo{author}{{Sawford}, B.~L.}
\newblock \bibinfo{title}{{Reynolds number effects in Lagrangian stochastic models of turbulent dispersion}}.
\newblock \emph{\bibinfo{journal}{Phys. Fluids A}} \textbf{\bibinfo{volume}{3}}, \bibinfo{pages}{1577--1586} (\bibinfo{year}{1991}).

\bibitem{frisch1995turbulence}
\bibinfo{author}{Frisch, U.}
\newblock \emph{\bibinfo{title}{Turbulence: the legacy of {A}. {N}. {K}olmogorov}} (\bibinfo{publisher}{Cambridge University Press}, \bibinfo{year}{1995}).

\bibitem{alexakisbiferale}
\bibinfo{author}{Alexakis, A.} \& \bibinfo{author}{Biferale, L.}
\newblock \bibinfo{title}{Cascades and transitions in turbulent flows}.
\newblock \emph{\bibinfo{journal}{Phys. Rep.}} \textbf{\bibinfo{volume}{767-769}}, \bibinfo{pages}{1--101} (\bibinfo{year}{2018}).

\bibitem{hinsberg2012}
\bibinfo{author}{van Hinsberg, M. A.~T.}, \bibinfo{author}{Thije~Boonkkamp, J. H.~M.}, \bibinfo{author}{Toschi, F.} \& \bibinfo{author}{Clercx, H. J.~H.}
\newblock \bibinfo{title}{On the efficiency and accuracy of interpolation methods for spectral codes}.
\newblock \emph{\bibinfo{journal}{SIAM J. Sci. Comput.}} \textbf{\bibinfo{volume}{34}}, \bibinfo{pages}{B479--B498} (\bibinfo{year}{2012}).

\bibitem{TURB-Rot}
\bibinfo{author}{Biferale, L.}, \bibinfo{author}{Bonaccorso, F.}, \bibinfo{author}{Buzzicotti, M.} \& \bibinfo{author}{Leoni, P. C.~D.}
\newblock \bibinfo{title}{{TURB-R}ot: a large database of 3{D} and 2{D} snapshots from turbulent rotating flows.} \bibinfo{note}{Preprint at \url{https://arxiv.org/abs/2006.07469} (2020)}.

\bibitem{TURB-Lagr}
\bibinfo{author}{Biferale, L.}, \bibinfo{author}{Bonaccorso, F.}, \bibinfo{author}{Buzzicotti, M.} \& \bibinfo{author}{Calascibetta, C.}
\newblock \bibinfo{title}{{TURB-L}agr: a database of 3{D} {Lagrangian} trajectories in homogeneous and isotropic turbulence.} \bibinfo{note}{Preprint at \url{https://arxiv.org/abs/2303.08662} (2023)}.

\bibitem{TURB-Hel}
\bibinfo{author}{Biferale, L.}, \bibinfo{author}{Bonaccorso, F.}, \bibinfo{author}{Linkmann, M.} \& \bibinfo{author}{Capocci, D.}
\newblock \bibinfo{title}{{TURB-H}el: an open-access database of helically forced homogeneous and isotropic turbulence.} \bibinfo{note}{Preprint at \url{https://arxiv.org/abs/2404.07653} (2024)}.

\bibitem{TURB-MHD}
\bibinfo{author}{Capocci, D.}, \bibinfo{author}{Biferale, L.}, \bibinfo{author}{Bonaccorso, F.} \& \bibinfo{author}{Linkmann, M.}
\newblock \bibinfo{title}{{TURB-MHD}: an open-access database of forced homogeneous magnetohydrodynamic turbulence}.
\newblock \bibinfo{note}{Preprint at \url{https://arxiv.org/abs/2504.10755} (2025)}.

\bibitem{heinonen2025lowinfo}
\bibinfo{author}{Heinonen, R.~A.}, \bibinfo{author}{Biferale, L.}, \bibinfo{author}{Celani, A.} \& \bibinfo{author}{Vergassola, M.}
\newblock \bibinfo{title}{Optimal trajectories for bayesian olfactory search in turbulent flows: The low information limit and beyond}.
\newblock \emph{\bibinfo{journal}{Phys. Rev. Fluids}} \textbf{\bibinfo{volume}{10}}, \bibinfo{pages}{044601} (\bibinfo{year}{2025}).

\bibitem{Arn_odo_2008}
\bibinfo{author}{Arn{\`{e}}odo, A.} \emph{et~al.}
\newblock \bibinfo{title}{Universal intermittent properties of particle trajectories in highly turbulent flows}.
\newblock \emph{\bibinfo{journal}{Phys. Rev. Lett.}} \textbf{\bibinfo{volume}{100}}, \bibinfo{pages}{254504} (\bibinfo{year}{2008}).

\bibitem{biferale2008lagrangianSF}
\bibinfo{author}{Biferale, L.} \emph{et~al.}
\newblock \bibinfo{title}{Lagrangian structure functions in turbulence: {A} quantitative comparison between experiment and direct numerical simulation}.
\newblock \emph{\bibinfo{journal}{Phys. Fluids}} \textbf{\bibinfo{volume}{20}}, \bibinfo{pages}{065103} (\bibinfo{year}{2008}).

\bibitem{benzi1993extended}
\bibinfo{author}{Benzi, R.} \emph{et~al.}
\newblock \bibinfo{title}{Extended self-similarity in turbulent flows}.
\newblock \emph{\bibinfo{journal}{Phys. Rev. E}} \textbf{\bibinfo{volume}{48}}, \bibinfo{pages}{R29--R32} (\bibinfo{year}{1993}).

\bibitem{EXP1}
\bibinfo{author}{Berg, J.}, \bibinfo{author}{L\"{u}thi, B.}, \bibinfo{author}{Mann, J.} \& \bibinfo{author}{Ott, S.}
\newblock \bibinfo{title}{Backwards and forwards relative dispersion in turbulent flow: {A}n experimental investigation}.
\newblock \emph{\bibinfo{journal}{Phys. Rev. E}} \textbf{\bibinfo{volume}{74}}, \bibinfo{pages}{016304} (\bibinfo{year}{2006}).

\bibitem{EXP2}
\bibinfo{author}{Xu, H.}, \bibinfo{author}{Bourgoin, M.}, \bibinfo{author}{Ouellette, N.~T.} \& \bibinfo{author}{Bodenschatz, E.}
\newblock \bibinfo{title}{High order {L}agrangian velocity statistics in turbulence}.
\newblock \emph{\bibinfo{journal}{Phys. Rev. Lett.}} \textbf{\bibinfo{volume}{96}}, \bibinfo{pages}{024503} (\bibinfo{year}{2006}).

\bibitem{EXP3}
\bibinfo{author}{Mordant, N.}, \bibinfo{author}{Metz, P.}, \bibinfo{author}{Michel, O.} \& \bibinfo{author}{Pinton, J.-F.}
\newblock \bibinfo{title}{Measurement of {L}agrangian velocity in fully developed turbulence}.
\newblock \emph{\bibinfo{journal}{Phys. Rev. Lett.}} \textbf{\bibinfo{volume}{87}}, \bibinfo{pages}{214501} (\bibinfo{year}{2001}).

\bibitem{DNS1}
\bibinfo{author}{Mordant, N.}, \bibinfo{author}{L\'{e}v\^{e}que, E.} \& \bibinfo{author}{Pinton, J.-F.}
\newblock \bibinfo{title}{Experimental and numerical study of the {L}agrangian dynamics of high {R}eynolds turbulence}.
\newblock \emph{\bibinfo{journal}{New J. Phys.}} \textbf{\bibinfo{volume}{6}}, \bibinfo{pages}{116} (\bibinfo{year}{2004}).

\bibitem{DNS2}
\bibinfo{author}{Homann, H.}, \bibinfo{author}{Grauer, R.}, \bibinfo{author}{Busse, A.} \& \bibinfo{author}{M\"{u}ller, W.~C.}
\newblock \bibinfo{title}{{L}agrangian statistics of {N}avier–{S}tokes and {MHD} turbulence}.
\newblock \emph{\bibinfo{journal}{J. Plasma Phys.}} \textbf{\bibinfo{volume}{73}}, \bibinfo{pages}{821–830} (\bibinfo{year}{2007}).

\bibitem{DNS3}
\bibinfo{author}{Biferale, L.}, \bibinfo{author}{Boffetta, G.}, \bibinfo{author}{Celani, A.}, \bibinfo{author}{Lanotte, A.} \& \bibinfo{author}{Toschi, F.}
\newblock \bibinfo{title}{Particle trapping in three-dimensional fully developed turbulence}.
\newblock \emph{\bibinfo{journal}{Phys. Fluids}} \textbf{\bibinfo{volume}{17}}, \bibinfo{pages}{021701} (\bibinfo{year}{2004}).

\bibitem{DNS4}
\bibinfo{author}{Fisher, R.~T.} \emph{et~al.}
\newblock \bibinfo{title}{Terascale turbulence computation using the {FLASH3} application framework on the {IBM B}lue {G}ene/{L} system}.
\newblock \emph{\bibinfo{journal}{IBM J. Res. Dev.}} \textbf{\bibinfo{volume}{52}}, \bibinfo{pages}{127--136} (\bibinfo{year}{2008}).

\bibitem{DNS5}
\bibinfo{author}{Yeung, P.~K.}, \bibinfo{author}{Pope, S.~B.} \& \bibinfo{author}{Sawford, B.~L.}
\newblock \bibinfo{title}{Reynolds number dependence of {L}agrangian statistics in large numerical simulations of isotropic turbulence}.
\newblock \emph{\bibinfo{journal}{J. Turbul.}} \textbf{\bibinfo{volume}{7}}, \bibinfo{pages}{N58} (\bibinfo{year}{2006}).

\bibitem{3bd-clause}
\bibinfo{author}{Opensource.org}.
\newblock \bibinfo{title}{\url{https://opensource.org/license/bsd-3-clause}} (\bibinfo{year}{2025}).

\end{thebibliography}

\end{document}